\documentstyle[epsf]{mn}
\input{psfig}
\title{Properties of the hard X-ray radiation from Cygnus~X-1 and
1E1740.7-2942} 

\author[S. Kuznetsov et al.]
  {S.~Kuznetsov,$^{1,2,3}$\thanks{E-mail: skuznetsov@hea.iki.rssi.ru}
	M.~Gilfanov,$^{1,3}$
	E.~Churazov,$^{1,3}$
	R.~Sunyaev,$^{1,3}$
	I.~Korel,$^{1,2}$
  \newauthor
	N.~Khavenson,$^{1}$
	A.~Dyachkov,$^{1}$
	I.~Chulkov,$^{1}$
	J.~Ballet,$^{4}$
	P.~Laurent,$^{4}$
	M.~Vargas,$^{4}$
  \newauthor
	A.~Goldwurm,$^{4}$
	J.P.~Roques,$^{5}$
	E.~Jourdain,$^{5}$
	L.~Bouchet,$^{5}$
	V.~Borrel.$^{5}$\\
$^{1}${Space Research Institute, Russian Academy of Sciences,
Profsouznaya 84/32, Moscow 117810, Russia}\\
$^{2}${Moscow Physical-Technical Institute, Institutsky Lane 9,
Dolgoprudny, Moscow area 141700, Russia}\\
$^{3}${Max-Planck-Institut fur Astrophysik, Karl-Schwarzschild-Str. 1,
85740 Garching bei Munchen, Germany}\\
$^{4}${Service d'Astrophysique, DAPNIA/DSM, Bt 709, CEA Saclay,
91191 Gif-sur-Yvette Cedex, France}\\
$^{5}${Centre d'Etude  Spatiale des Rayonnements (CNRS/UPS) 9,
avenue du Colonel Roche, BP 4346, 31028 Toulouse Cedex, France}}
\date{Accepted  Received }
\pagerange{\pageref{firstpage}--\pageref{lastpage}}
\pubyear{1997}

\begin{document}
 
\label{firstpage}
 
\maketitle

%===========================================================================%
%%%%%%%%%%%%%%%%%%%%%%%%%%%%% A B S T R A C T %%%%%%%%%%%%%%%%%%%%%%%%%%%%%%%
%===========================================================================%

\begin{abstract}
The entire dataset of the GRANAT/SIGMA observations of Cyg X-1 and
1E1740.7-2942 in 1990--1994 was analyzed in order to search for correlations
between primary observational characteristics of the hard X-ray (40--200
keV) emission -- the hard X-ray luminosity $L_X$, the hardness of the spectrum
(quantified in terms of the best-fit thermal bremsstrahlung temperature
$kT$) and the {\em rms} of short-term flux variations.

Although no strict point-to-point correlations were detected certain general
tendencies are evident. It was found that for Cyg X--1 the spectral hardness
is in general positively correlated with the relative amplitude of short-term
variability. A correlation of a similar kind was found for the X-ray transient
GRO J0422+32 (X-ray Nova Persei 1992).

For both sources an approximate correlation between $kT$ and $L_X$ was found.
At low hard X-ray luminosity -- below $\sim 10^{37}$ erg/sec -- $kT$
increases with $L_X$. At higher luminosity the spectral hardness
depends weaker or does not depend at all on the hard X-ray luminosity. The
low luminosity end of these approximate correlations (low $kT$ and low
$rms$) corresponds to extended episodes of very low hard X-ray flux which
occurred during SIGMA observations.

\end{abstract}


%===========================================================================%
%%%%%%%%%%%%%%%%%%%%%%%% I N T R O D U C T I O N %%%%%%%%%%%%%%%%%%%%%%%%%%%%
%===========================================================================%
\section{Introduction}

%**************** C Y G N U S   X - 1 ***************************************

Cygnus~X-1 was the first dynamically proven black hole candidate in the
Galaxy and it's X--ray properties were studied on a number of occasions for
the last two decades.  The source exhibits flux variations on all
time-scales from years to milliseconds. Two distinct spectral states of the
source were identified \cite{tan72,holt76,oga82}: the ``Low State'' (LS) and
the ``High State'' (HS). In the LS emission from the source in the X--ray
domain is dominated by a hard spectral component -- most likely Comptonized
radiation -- observed up to several hundred keV. The HS spectrum is much
softer and is dominated by a soft spectral component.  Most of the time
($\sim 90\%$) the source was found in the LS (e.g. Liang \& Nolan 1984).
Basing on the observations of HEAO-3 Ling et al. (1987) proposed to
distinguish three sub-states of the LS -- the so called $\gamma$--states
characterized by different intensity and spectral properties of the X- and
$\gamma$-ray emission.

\begin{figure*}
\includegraphics{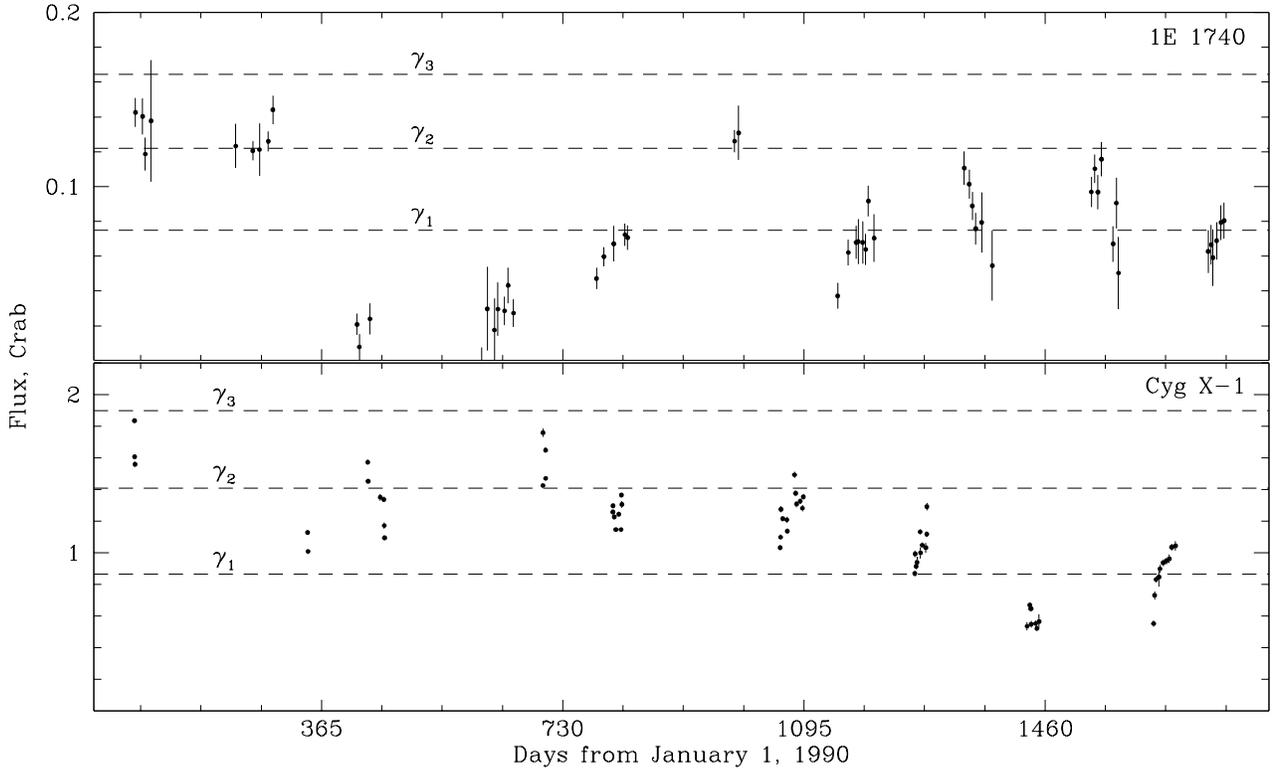}
	\vspace*{300pt}
	\caption[Figure1.ps]{ 	
The 40--150 keV light curves of Cyg
X-1 (bottom) and 1E1740.7--2942 (top).  Each data point represents the
average over $\approx 2-8$ hours (Cyg X-1) and $\sim 50$ hours
(1E1740.7--2942). The date Jan.1,1990 corresponds to MJD 47892. Approximate
flux levels corresponding to the three $\gamma$--states of Cyg X-1 are
shown on both panels.
The $\gamma$--states in the top panel (1E1740.7-2942) were rescaled to the
Galactic Center distance.}
	\label{Fig1}
\end{figure*}

%******************* 1 E 1 7 4 0 . 7 - 2 9 4 2 ******************************

The source 1E1740.7-2942 (hereafter 1E~1740) located 50 arcmin apart from
the Galactic Center is the hardest X-ray source in this region. On the basis
of it's X-ray properties it was suggested, that this source too contains a
black hole \cite{rs91a}. Its spectral shape and X-ray luminosity are quite
similar to that of Cygnus~X-1 in its ``nominal'' $\gamma_2$\ state
\cite{rs91b}.

In this letter we report the results of a search for correlations between
the primary observational characteristics of the hard X-ray emission during
the low state of these two sources: the hard X-ray luminosity, the hardness
of the spectrum and the amplitude of short-term flux variations. Preliminary
results of this analysis were presented earlier by Kuznetsov et al. (1995)
and Ballet et al.  (1996). Similar results were also obtained for Cyg X-1 by
Crary et al. (1996) and Phlips et al. (1996) basing on the BATSE and OSSE data.

%===========================================================================%
%%%%%%%%%%%%% I N S T R U M E N T S  AND  O B S E R V A T I O N S  %%%%%%%%%%
%===========================================================================%

\section{Instruments and Observations}

The SIGMA hard X-ray/soft $\gamma$-ray telescope is one of the two main
imaging instruments on board the GRANAT observatory.  A detailed description
of the telescope is given by Paul et al.(1991).

The observations of Cygnus~X-1 have been carried out twice a year: during
December in 1990--1993, March in 1990--1992 and June in 1993--1994.  Results
of the first Cyg~X-1 observations with GRANAT/SIGMA were presented by
Salotti et al. (1992).  1E~1740 was observed mostly during each Spring
and Fall in 1990-1994 (sometimes including a part of February and August).
The total exposure time taken by the SIGMA telescope during 33 pointed
observations of Cyg X-1 was $\sim$600 hours. The 1E~1740 source was
within the SIGMA field of view during 124 observations in the course of 10
Galactic Center surveys performed by GRANAT in 1990-1994 with total exposure
time of $\sim$2100 hours.

The long-term light curve of Cygnus~X-1 (Fig.1) recorded by SIGMA shows
variations of the 40-150 keV flux by a factor of $\sim$4.  During 1990 --
mid 1993 \cite{vano} the source was typically detected near the $\gamma_2$
intensity level with a maximum flux of $\sim$1.8 Crab (corresponding to
$\sim \gamma _3$ level) observed on 23-24 March 1990.  In Dec 1993 and in
the first observations in June 1994 the minimal flux was detected, $\sim
0.5$ Crab (corresponding to $\sim 0.6\gamma_1$).  According to BATSE data
\cite{crary} having much better time coverage these two observational sets
occurred during an extended low hard X-ray flux episode which started in
Sept.1993 and lasted for $\sim$ one year. The lowest flux from Cyg X-1
during this period was detected in Feb.1994 ($0.2\gamma_1$ -- Phlips et al.
1996). During almost all SIGMA observations considerable variability on the
hours--days time scale was detected -- by a factor of $\sim 1.5$.

The light curve of 1E~1740 recorded by SIGMA (Fig.1) shows a
qualitatively similar pattern \cite{cord}. The 40-150 keV flux from
1E~1740 changed by a factor of $\sim$10 on the time-scale of $\sim$\ 1/2
year with the minimal flux corresponding to the extended minimum observed
during 1991 \cite{chur93}.  During the first half of Fall 1991 observations
the source flux was below SIGMA detection limit (3$\sigma$ - $\sim$ 13
mCrab). Variability by a factor of 1.5 on a days-weeks timescale was detected
during most of the observational sets. The relative weakness of
the source didn't allow to study flux variations on the hours time scale.

%===========================================================================%
%%%%%%%%%%%%%%%%%%%%%%%%% D A T A   A N A L Y S I S %%%%%%%%%%%%%%%%%%%%%%%%%
%===========================================================================%
\section{The Data Analysis}

\begin{figure}
	\psfig{figure=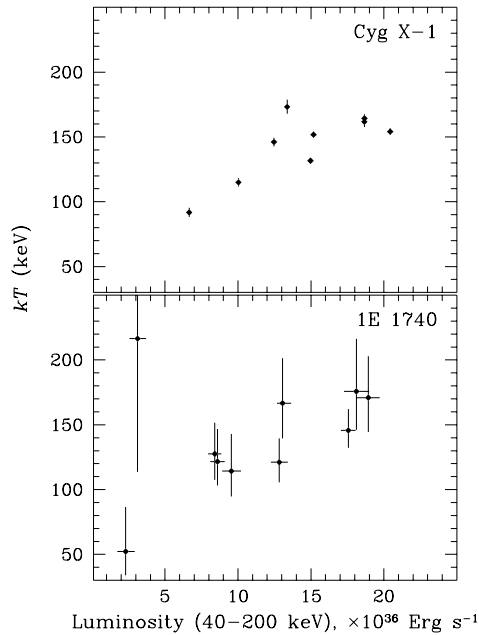,height=8.8truecm}
	\caption[Figure2.ps]{
The best-fit bremsstrahlung temperature plotted against the hard X-ray
luminosity (40-200 keV) for Cyg X-1 (upper panel) and 1E~1740.7--2942 (lower
panel). The data were averaged over 1 to 20 days of consecutive observations.} 
\label{Fig2}
\end{figure}

Only observations performed under nominal background and instrument
conditions were used for the analysis. Since the timing analysis is more
sensitive to nonstandard background conditions, whenever timing analysis was
impossible or ambiguous the dataset was excluded from the spectral analysis
too.

In order to quantify the hardness of the source emission in the 40-200 keV
energy range the optically-thin thermal bremsstrahlung model \cite{kel} was
chosen. Although having no direct physical relation to the origin
of the hard X-ray emission in compact sources, it provides a good
approximation to the observed spectra and characterizes the spectral shape
by a single parameter -- the best-fit temperature. The relative error
between the data and the spectral model is less than 10\% in the 40-200 keV
energy band.  The model was applied to the pulse-height spectra averaged as
described in the next two sections.  The 40--200 keV luminosity was
calculated using the best fit model and assuming a distance of 2.5 and 8.5
kpc for Cyg X-1 and 1E~1740 respectively.

\begin{figure}
	\psfig{figure=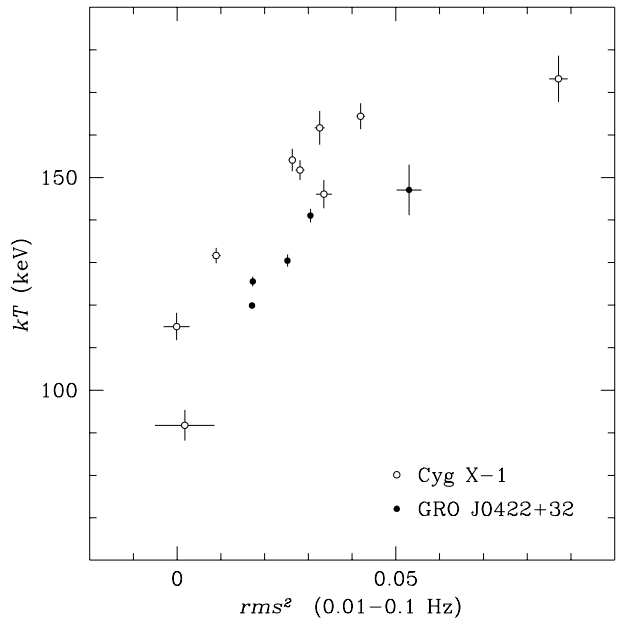,height=8.8truecm}
	\caption[Figure3.ps]{
The best-fit bremsstrahlung temperature plotted
against the $rms^{2}$\ of the short-term flux variations in the 0.01--0.1 Hz
frequency range. Open circles correspond to Cygnus~X-1, filled circles -- to
X-ray Nova Persei 1992 (GRO J0422+32). The data were averaged in the same
way as in Fig. 2.} 
\label{Fig3}
\end{figure}

For the timing analysis the 4 s resolution 40--150 keV data (count rate from
the entire detector) were used. The power density spectrum was obtained for
each individual spectral image (SI) exposure using the standard timing
analysis technique \cite{klis} and then converted to units of relative {\em
rms} using the source intensity measured in the image of the same SI
exposure.  The values of relative {\em rms} were further averaged as
described in the next two sections. The aperiodic variability of the source
was quantified by the fractional {\it rms} of the flux variations in the
0.01-0.1 {\it Hz} frequency band. This range corresponds to the flat part in
the power density spectrum of Cygnus~X-1 and represents its most variable
part. The timing analysis was not performed for 1E~1740 since this source is
significantly fainter and located in the crowded Galactic Center region.

It should be noted that opposite to the spectral analysis which
utilizes the imaging capability of the SIGMA telescope the 4 s
resolution data used for the timing analysis does not possess spatial
resolution and the count rate from the entire detector was analyzed.
Numerous tests confirmed that under standard background conditions
possible contamination of the Cyg X-1 power density spectrum by
background events in the 0.01-0.1 {\it Hz} frequency range can be
neglected.

\section{Results.}

\begin{figure}
	\psfig{figure=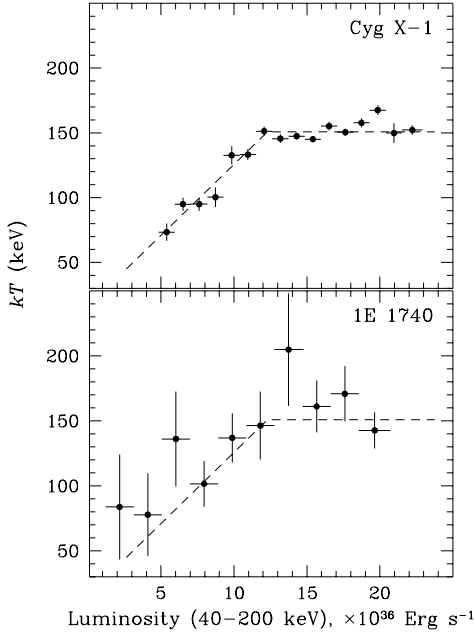,height=8.8truecm}
	\caption[Figure4.ps]{
The best-fit bremsstrahlung temperature plotted against the hard X-ray
luminosity (40-200 keV) for Cyg X-1 (upper panel) and 1E1740.7--2942 (lower
panel). The data were averaged according to the X-ray intensity as described
in section 4.2. The approximation by a linear trend (with offset) and
constant to Cygnus~X-1 data is shown by a dashed line in both panels.}
\label{Fig4}
\end{figure}

As it was described above the light curves of both sources have a complex
structure with short term (time scales of days to weeks) variations
superimposed on long term (time scale of years) intensity changes of
generally larger relative amplitude. The SIGMA observations provided on one
hand rather sparse time coverage -- especially for Cyg X-1, and, on the
other, a limited time resolution restricted by the instrument time
resolution (several hours for spectral information) and, especially for
1E~1740, by the accuracy of the spectral and variability parameters
estimation. The latter leads to the necessity of further grouping of the data.
In order to verify possible effects of the data averaging two grouping
methods were applied to the data as described below.

\subsection{Grouping by observational sets.}

In order to study the source behavior on the time-scale of months we have
averaged the data acquired during individual observational sets.  The typical
time span for each data point was $\sim$1 day to $\sim$1 month. Results are
shown in Fig. 2 and 3.
  
\begin{figure}
	\psfig{figure=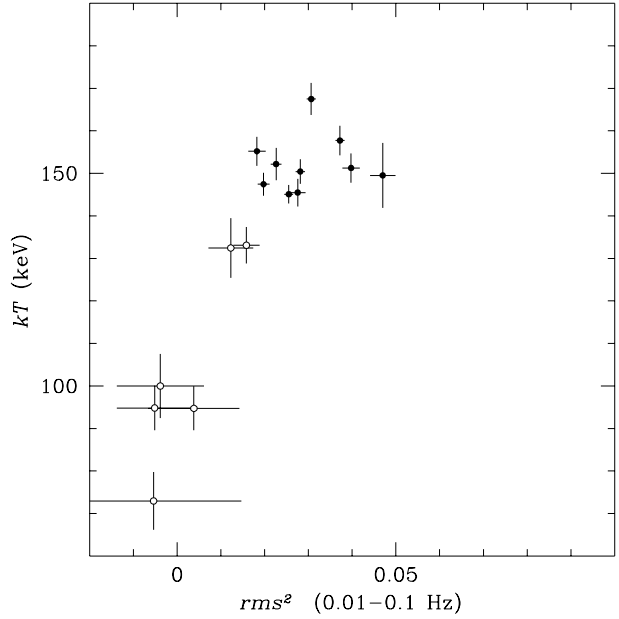,height=8.8truecm}
	\caption[Figure5.ps]{
The best-fit bremsstrahlung temperature plotted
against the $rms^{2}$\ of the short-term flux variations (0.01--0.1 Hz)  for
Cyg~X-1.  The data were averaged in the same way as in Fig. 4.
The open circles correspond to points with luminosity below 1.2 $\times$\
10$^{37}$ erg/s, filled circles -- above that luminosity level (see Fig. 4).
}
\label{Fig5}
\end{figure}

The relation between spectral hardness and luminosity is shown in Fig. 2. A
trend is present in the data for both sources -- $kT$ generally increases as
the hard X-ray luminosity increases.

The Fig. 3 shows the dependence of the best-fit bremsstrahlung temperature
upon the {\it rms} of the short-term flux variations for Cyg~X-1.  There is a
clear correlation between the spectrum hardness and the amplitude of the
low-frequency noise (0.01-0.1 {\it Hz}) in the 40-150 keV energy band.  It
is very important that the same behavior is found in X-ray Nova Persei. In
order to qualitatively illustrate the similarities in the properties of the
hard X-ray emission of both sources the GRO J0422+32 data are shown in Fig.
3.

Apart from the two low flux - low $rms$\ points, there was no obvious
correlation between rms and luminosity. This is a difference with Nova
Persei where the evolution was smooth and apparently controlled by a single
parameter.

\subsection{Grouping according to intensity.}

The original data were regrouped according to source intensity in the
following way. The entire range of the 40-200 keV flux variations was
divided into a number of bins of equal width.  The mean energy and power
density spectra corresponding to each intensity bin were calculated by
averaging over all individual datasets with intensity falling 
into that bin.

For Cyg X-1 16 intensity bins were chosen covering the 1.9 to 6.9
$\times10^{-2}$ cnt/s/cm$^{2}$\ (0.5-1.8 Crab) intensity range.  The
regrouping procedure was applied to the data of individual SI exposures (4-8
hours long - the highest time resolution providing spectral information)
each exposure being treated as a separate dataset.  In the case of 1E~1740,
having 5 to 20 times lower signal to noise ratio the data averaged over each
single observation (comprised of 1-6 SI exposures with total duration of
4-34 hours) were treated as individual datasets to be regrouped. The
intensity range 0.3 to 5.6 $\times10^{-3}$ cnt/s/cm$^{2}$\ (8-150 mCrab) was
divided into 10 intensity bins.

The dependence of the best-fit bremsstrahlung temperature upon the 40-200
keV luminosity for both sources is shown in Fig. 4. The similarity in the
behavior of the two sources is apparent.  Cyg X-1 data can be represented by
a linear trend (with offset) at low flux and constant temperature at high
flux (shown as a dashed line). The same curve describes quite well the data
for 1E~1740. The ``breaks'' in Figure 4 correspond to the same value of the
luminosity. The kT-$rms^{2}$\ dependence for Cyg X-1 is shown in Figure 5.

The statistical significance of the correlations shown in Fig.4 was
estimated using Nonparametric or Rank Correlation test. The values of
Spearman Rank-Order correlation coefficient giving probability that these
correlations resulted from statistical fluctuations, are $\sim$\ 10$^{-5}$
($r_{s}$ = 0.874, 16 bins) and $\sim$\ 4 $\times$\ 10$^{-3}$\
($r_{s}$ = 0.818, 10 bins) for Cyg X-1 and 1E~1740 respectively.

It should be noted, that shown in Figure 4 are parameters derived from
averaged spectra for each intensity bin and the error bars are statistical
only. The analysis of individual datasets, corresponding to a given
intensity bin, performed with the more statistically significant Cyg X-1
data, revealed considerable dispersion of the best-fit parameters above the
level of statistical fluctuations. This dispersion is of the order of
$\sim$15\% of the values of the best-fit temperature shown in Figure 4.  It
is especially large in the $(0.8-1.2)\times 10^{37}$ erg/sec luminosity
range. Therefore, the dependence shown in Fig. 4 is not a point-to-point
correlation, but rather represents some averaged pattern of the behavior of
the parameters.

\begin{figure}
	\psfig{figure=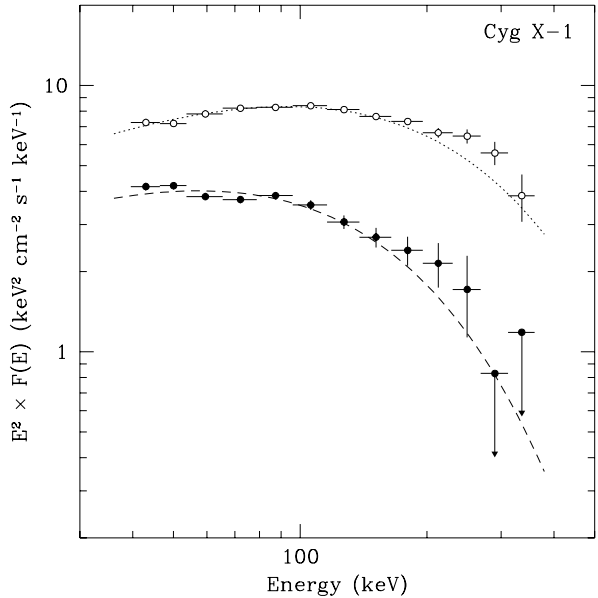,height=8.8truecm}
	\caption[Figure6.ps]{The spectra of Cyg~X-1 in the state with 
high (hollow circles) and low (filled circles) 40-200 keV flux.
The high flux spectrum corresponds to the horizontal part of the $L_X-kT$\
dependence  shown in Fig. 4 and was averaged over March 21-27, 1992
observations. The low flux spectrum (low luminosity end of the dependence in
Fig. 4) was averaged over December 4-13, 1993 observations.
Best-fit approximation by optically-thin thermal bremsstrahlung model is
shown for both spectra. Points below 1$\sigma$\ level are replaced with
1$\sigma$\ upper limit.
}
\label{Fig6}
\end{figure}

Figure 6 shows typical source spectra at the horizontal (high 40-200 keV
luminosity) and the increasing (low 40-200 keV luminosity) parts of the
$L_X-kT$\ dependence in Fig. 4.   The best-fit bremsstrahlung
temperatures are: $kT = 148 \pm 2$\ keV and $kT = 91
\pm 4$ keV respectively.

The spectra of Cyg X-1 and 1E~1740 averaged over all observations when the
40-200 keV luminosity was above $1.2 \times 10^{37}$\ erg/sec (i.e. on the
flat part of the $L_X-kT$\ dependence in Fig. 4) are shown in Fig.7. The
spectrum of 1E~1740 was scaled from a distance of 8.5 kpc to 2.5 kpc for
comparison with Cyg X-1.  Similarity of the spectra of both sources is
apparent.

\begin{figure}
	\psfig{figure=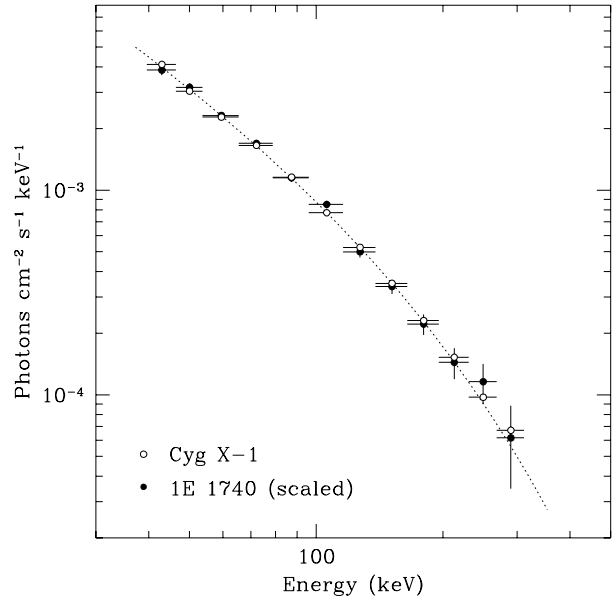,height=8.8truecm}
	\caption[Figure7.ps]{The spectra of Cyg~X-1 and 1E1740.7-2942 averaged
over all observations when the 40-200 keV luminosity was above $1.2 \times
10^{37}$\ erg/sec (corresponding to the break in Fig. 4).  The 1E1740.7-2942
spectrum was scaled to  Cyg~X-1 distance (2.5 kpc).
The best-fit thermal bremsstrahlung spectrum to Cyg~X-1 data is
shown by dotted line. 
}
\label{Fig7}
\end{figure}

%===========================================================================%
%%%%%%%%%%%%%%%%%%%%%%%%% D I S C U S S I O N  %%%%%%%%%%%%%%%%%%%%%%%%%%%%%%
%===========================================================================%
\section{Discussion}

Monitoring the black hole candidates Cyg X-1 and 1E~1740 with the SIGMA
telescope revealed two general tendencies in their behavior:
\begin{enumerate}
\item For both sources the hardness of the spectrum generally increases as
hard X-ray luminosity (40--200 keV) increases up to $\sim$\ 10$^{37}$\ erg/s
and is nearly constant at higher luminosity.
\item In the case of Cyg X--1 the relative amplitude of  short term
aperiodic variability generally increases with spectral hardness.
\end{enumerate}

Lower values of $kT$ and $rms$ associated with a lower luminosity in the
hard X-ray band might correspond to a transition of the source to the high
or very high spectral state. The hard spectral component in the very high
state of black hole candidates is generally steeper and less variable on
short time scales than in the low state (e.g. Miyamoto et al. 1991).  It was
suggested (e.g. Miyamoto et al. 1991) that these differences reflect a
different origin of the hard spectral component in the low and (very) high
spectral states. In the accretion disk model considered by Chakrabarti and
Titarchuk (1995) the hard spectral component is generated via Comptonization
of soft photons from the optically thick Keplerian disk either in a hot
optically slim post shock region at lower accretion rate (low spectral
state) or in a convergent flow at higher accretion rate (high spectral
state). These two mechanisms result in hard spectral components having
different spectral and variability properties. At intermediate values of the
disk accretion rate both mechanisms could contribute to the observed
emission provided that the halo accretion rate is sufficiently high. Thus,
with an increase of the accretion rate a continuous change of spectral and
short-term variability properties would occur as a result of the change of
the relative contributions of these two mechanisms.  In this scenario the
decrease of the hard X-ray luminosity does not reflect the change of the
accretion rate which is increasing.

Alternatively, the observed change of the spectral hardness might not be
connected with a transition to another spectral state but reflect subtle
changes of conditions in the Comptonization region either related or not
to a change of the accretion rate.  Whatever the origin of these changes 
the pattern in Fig. 3 and 5 hints at a close relation between the hardness of
the emergent spectrum and the level of the short-term flux variations.

The relation between the spectral hardness and the level of aperiodic
variability might be a very general property of the mechanism responsible
for the hard X-ray component in compact sources. This is supported by
comparison with other Galactic black hole candidates. A dependence of the
spectral hardness upon the hard X-ray luminosity opposite to that of Fig.4
was observed for several X-ray Novae -- black hole candidates (e.g. Nova Per
1992, Vikhlinin et al. 1995, and Nova Oph 1993, Gilfanov et al. 1993) -- for
these sources the spectral hardness is anti-correlated with
luminosity.  Nevertheless the relation between the relative rms of aperiodic
variability and the spectral hardness for X-ray Nova Per 1992 is very
similar to that observed for Cyg X-1 (Fig 3).

Finally it's worth noting that the correlations discussed above become apparent
mainly due to the data points acquired during the extended low intensity
episodes observed for both sources. They become much less evident when these
data are excluded from consideration. The Cyg X-1 data separated into two
parts -- the data acquired before and after December 1993 -- splits into two
independent branches on the $kT-L_X$ diagram overlapping in luminosity
-- nearly flat ($kT$ does not depend on $L_X$) and increasing ($kT$ is
positively correlated with $L_X$) (see Ballet et al., 1996 for details).
This might indicate the existence of two modes of the source behavior. With
the presently available dataset having rather sparse time coverage it is
possible neither to confirm nor reject this possibility.

\section*{Acknowledgments}
The IKI co-authors would like to acknowledge partial support of this work by
INTAS grant 93-3364 and RBRF grant 96-02-18588-A. S.Kuznetsov was also
partially supported by the ISSEP grants S96-207 and A97-2301.

\label{lastpage}

\end{document}